# NUMERICAL SOLUTION OF THE MODEL FOR HIV INFECTION OF CD4$^+$T CELLS BY USING MULTISTAGE DIFFERENTIAL TRANSFORM METHOD.

AUTHORS: KOLEBAJE OLUSOLA T. [a][*], OYEWANDE E.O.[a] AND MAJOLAGBE B.S.[b]

[a] Department of Physics, University of Ibadan, Ibadan, Nigeria.

[b] National Agency for Science and Engineering Infrastructure, Federal Ministry of Science and Technology, Ido Ilesa, Nigeria.

[*] Corresponding Author

POSTAL ADDRESS: Department of Physics, University of Ibadan, Ibadan, Nigeria.

EMAIL ADDRESS: olusolakolebaje2008@gmail.com

TELEPHONE NO: +2347038680315

## ABSTRACT

The Multistage Differential Transform Method (MDTM) is employed to solve the model for HIV infection of CD4$^+$T cells. Comparing the numerical results to those obtained by the classical fourth order Runge-Kutta method showed the preciseness and efficacy of the multistep differential transform method. The study shows that the method is a powerful and promising tool for solving coupled systems of differential equations.

## 1.0 INTRODUCTION

Human Immunodeficiency Virus (HIV) is a retrovirus that causes Acquired Immunodeficiency Syndrome (AIDS) which is characterized by the progressive failure of the human immune system. AIDS is a major public health, sociological and economic concern and a cause of deaths in many parts of Africa. Africa is estimated to be home to 69% of all people living with HIV and 72% of all AIDS death in 2009 [1].

HIV primarily infects a class of white blood cells called CD4$^+$T cells and this selective depletion of CD4$^+$T cells which plays a central role in immune regulation serves as a clinical indicator for measuring the progression of HIV infection. Mathematical models have played an important role in understanding the dynamics of this infectious disease [1-6]. One of such models for the HIV infection of CD4$^+$T cells is given by the following system of differential equations [4]:

$$\frac{dT}{dt} = s - \alpha T + rT\left(1 - \frac{T+I}{T_{max}}\right) - kVT$$

$$\frac{dI}{dt} = kVT - \beta I \qquad (1)$$

$$\frac{dV}{dt} = N\beta I - \gamma V$$

With the initial conditions:

$T(t) = 0.1 \; cells/mm^3$, $I(t) = 0 \; cells/mm^3$ and $V(t) = 0.1 \; virions/mm^3$

Where $T(t)$, $I(t)$ and $V(t)$ denote the concentration of CD4$^+$T cells, the concentration of infected CD4$^+$T cells by the HIV viruses, and free HIV virus particles respectively.

**Table 1: Parameter description and values.**

| Parameter | Description | Value | Units | Source |
|---|---|---|---|---|
| $s$ | CD4$^+$T cells constant source | 0.1 | $cells/mm^3/day$ | [6] |
| $\alpha$ | CD4$^+$T cells natural turn-over rate | 0.02 | $day^{-1}$ | " |
| $r$ | CD4$^+$T cells production rate through mitosis | 3 | $day^{-1}$ | " |
| $T_{max}$ | Maximum concentration of CD4$^+$T cells in the body | 1500 | $cells/mm^3$ | " |
| $k$ | viral activity rate | 0.0027 | $mm^3/cells/day$ | " |
| $\beta$ | death rate of infected cells | 0.3 | $day^{-1}$ | " |
| $N$ | production rate of virus particles | 10 | $virions/cell$ | " |
| $\gamma$ | viral production rate by busting CD4$^+$T cells | 2.4 | $day^{-1}$ | " |

The merit of the Differential Transform method (DTM) is that the method does not require discretization or perturbation and is easy to implement, while also greatly reducing the size of computational work to be done. However, the DTM does not give a satisfactory approximation for a large time $t$ as observed in [7], [8] where the convergence region of semi-analytic methods are said to be narrow. Since the DTM solutions blow out after a short time a multi-staging technique known as the Multistage Differential Transform Method (MDTM) is proposed.

The motivation of this paper is to apply the DTM and MDTM to the model of HIV infection of CD4$^+$T cells and test the accuracy of the method with the well known fourth order Runge-Kutta method (RK4). The computations in this paper were carried out with the computer algebraic system MATHEMATICA.

## 2.0 METHODOLOGY

### 2.1 DIFFERENTIAL TRANSFORM METHOD (DTM)

If the function $u(x,t)$ is analytic and differentiable continuously with respect to time $t$ and space $x$ in the domain of interest, then the function can be expanded in Taylor series about a point $t = t_0$ as;

$$u(x,t) = \sum_{k=0}^{\infty} \frac{1}{k!} \left[\frac{\partial^k u(x,t)}{\partial t^k}\right]_{t=0} (t-t_0)^k \qquad (2)$$

The differential transformation of $u(x,t)$ is defined as

$$U_k(x) = \frac{1}{k!} \left[\frac{\partial^k u(x,t)}{\partial t^k}\right]_{t=t_0} \qquad (3)$$

Where $U_k(x)$ known as the $t$-dimensional spectrum function is the transformation of the function $u(x,t)$. The inverse differential transform of $U_k(x)$ is defined as follows;

$$u(x,t) = \sum_{k=0}^{\infty} U_k(x)(t-t_0)^k \qquad (4)$$

When $t_0 = 0$, then equations (2) and (3) are expressed as

$$u(x,t) = \sum_{k=0}^{\infty} \frac{1}{k!} \left[\frac{\partial^k u(x,t)}{\partial t^k}\right]_{t=0} t^k \qquad (5)$$

$$U_k(x) = \frac{1}{k!} \left[\frac{\partial^k u(x,t)}{\partial t^k}\right]_{t=0} \qquad (6)$$

$$u(x,t) = \sum_{k=0}^{\infty} U_k(x) t^k \qquad (7)$$

The inverse transformation of the set of values $\{U_k(x)\}_{k=0}^n$ gives approximate solution as

$$\overline{u_n}(x,t) = \sum_{k=0}^{n} U_k(x) t^k \qquad (8)$$

Where $n$ is the order of the approximation. Therefore, the exact solution of the equation is given by;

$$u(x,t) = \lim_{n \to \infty} \overline{u_n}(x,t) \qquad (9)$$

The differential transformations of common functions following equation (6) are presented in Table 2.

**Table 2: Differential Transform Table**

| FUNCTIONAL FORM | TRANSFORMED FORM |
|---|---|
| $u(x,t)$ | $U_k(x) = \dfrac{1}{k!}\left[\dfrac{\partial^k u(x,t)}{\partial t^k}\right]_{t=0}$ |
| $u(x,t) \pm v(x,t)$ | $U_k(x) \pm V_k(x)$ |
| $\alpha u(x,t)$ | $\alpha U_k(x)$ ($\alpha$ is a constant) |
| $x^m t^n$ | $x^m \delta(k-n)$ |
| $x^m t^n u(x,t)$ | $x^m U(k-n)$ |
| $u(x,t)\, v(x,t)$ | $\sum_{r=0}^{k} U_r(x)\, V_{k-r}(x)$ |
| $\dfrac{\partial^r}{\partial t^r} u(x,t)$ | $\dfrac{(k+r)!}{k!} U_{k+r}(x)$ |
| $\dfrac{\partial}{\partial x} u(x,t)$ | $\dfrac{\partial}{\partial x} U_k(x)$ |
| Nonlinear Function $N(u(x,t)) = F(u(x,t))$ | $N_k = \dfrac{1}{k!}\left[\dfrac{\partial^k}{\partial t^k} F(U_0)\right]_{t=0}$ |

## 2.2 MULTISTAGE DIFFERENTIAL TRANSFORM METHOD (MDTM)

An efficient way of ensuring the validity of solutions to differential equations for large $t$ ($t \gg 0$) is by multi-staging the solution procedure to be employed. Let $[0, Z]$ be the interval over which the solutions to the differential equation (1) is to be determined. The solution interval $[0, Z]$ is divided into $N$ subintervals ($n = 1, 2, \ldots\ldots\ldots, N$) of equal step size given by $h = Z/N$ with the interval end points $t_n = nh$.

Initially, the DTM scheme is applied to obtain the approximate solutions of $T, I$ and $V$ of (1) over the interval $[0, t_1]$ by using the initial condition $T(0)$, $I(0)$ and $V(0)$ respectively. For obtaining the approximate solution of (1) over the next interval $[t_1, t_2]$, we take $T(t_1)$, $I(t_1)$ and $V(t_1)$ as the initial condition. Generally the scheme is repeated for any $n$ with the right endpoints $T(t_{m-1})$, $I(t_{m-1})$ and $V(t_{m-1})$ at the previous interval being used as the initial condition for the interval $[t_{m-1}, t_m]$.

## 3.0 APPLICATION

By using the fundamental operations of differential transformation method (5)-(9) and Table 1, the transformed form of the model of HIV infection of CD4+T cells is given below:

$$\frac{(n+1)!}{n!}T_{n+1} = s + (r-\alpha)T_n - \frac{r}{T_{max}}\sum_{a=0}^{n}T_a T_{n-a} - \frac{r}{T_{max}}\sum_{a=0}^{n}T_a I_{n-a} - k\sum_{a=0}^{n}V_a T_{n-a}$$

$$\frac{(n+1)!}{n!}I_{n+1} = k\sum_{a=0}^{n}V_a T_{n-a} - \beta I_n \qquad (10)$$

$$\frac{(n+1)!}{n!}V_{n+1} = N\beta I_n - \gamma V_n$$

The system is solved with the initial condition $T_0 = 0.1$, $I_0 = 0$ and $V_0 = 0.1$. The recurrence relation (10) was evaluated with the aid of Mathematica to obtain the solution up to an 11 terms approximation for the time range [0, 18] with a time step size 0.01. MDTM is implemented by dividing the solution interval $[0, 18]$ into 150 subintervals ($n = 1, 2, \ldots \ldots, 150$) of equal step size given by $h = 0.12$.

## 4.0 RESULTS AND DISCUSSION

The model of HIV infection of CD4+T cells was solved using the DTM and MDTM implemented on Mathematica, the 11 terms approximate DTM solution was obtained as:

$$T = 0.1 + 0.397953t + 0.642849t^2 + 0.671708t^3 + 0.525096t^4 + 0.332537t^5 + 0.181377t^6 + 0.091093t^7 + 0.046108t^8 + 0.026147t^9 + 0.017642t^{10}$$

$$I = 0.000027t + 0.000017t^2 - 3.90515e - 6t^3 + 3.31152e - 6t^4 - 9.56541e - 7t^5 + 4.83119e - 7t^6 - 4.95705e - 8t^7 + 7.92420e - 8t^8 + 2.90529e - 8t^9 + 3.04438e - 8t^{10}$$

$$V = 0.1 - 0.24t + 0.288041t^2 - 0.230415t^3 + 0.138246t^4 - 0.066356t^5 + 0.026542t^6$$
$$- 0.009100t^7 + 0.002730t^8 - 0.000728t^9 + 0.000175t^{10}$$

**Table 3: Absolute differences between 11-term DTM and 11-term MDTM with RK4 solutions** $(\Delta t = 0.01)$.

|   | $|DTM_{0.01} - RK4_{0.01}|$ | | | $|MDTM_{0.01} - RK4_{0.01}|$ | | |
|---|---|---|---|---|---|---|
| $t$ | $\Delta T$ | $\Delta I$ | $\Delta V$ | $\Delta T$ | $\Delta I$ | $\Delta V$ |
| 0.04 | 8.554E-05 | 2.809E-10 | 1.218E-09 | 8.554E-05 | 2.809E-10 | 1.218E-09 |
| 0.27 | 5.868E-03 | 7.598E-08 | 1.368E-08 | 4.731E-05 | 6.705E-11 | 5.632E-11 |
| 0.88 | 2.441E-01 | 2.197E-06 | 8.950E-06 | 8.554E-05 | 3.751E-11 | 1.798E-11 |
| 1.60 | 7.744E+00 | 2.254E-05 | 5.067E-03 | 8.552E-05 | 6.888E-12 | 3.130E-12 |
| 5.00 | 2.513E+05 | 3.677E-01 | 9.109E+02 | 7.907E-05 | 3.526E-11 | 1.510E-11 |
| 8.80 | 5.954E+07 | 1.178E+01 | 3.266E+05 | 7.911E-05 | 1.976E-07 | 9.050E-08 |
| 10.00 | 2.083E+08 | 4.873E+02 | 1.222E+06 | 7.937E-05 | 1.983E-06 | 6.885E-07 |
| 12.40 | 1.728E+09 | 2.085E+03 | 1.117E+07 | 8.056E-05 | 3.314E-06 | 2.940E-07 |
| 14.80 | 9.908E+09 | 1.614E+04 | 6.840E+07 | 8.301E-05 | 1.690E-06 | 1.592E-07 |
| 16.00 | 2.142E+10 | 3.553E+04 | 1.517E+08 | 8.374E-05 | 1.197E-06 | 1.101E-07 |
| 18.00 | 6.875E+10 | 1.151E+05 | 5.047E+08 | 8.399E-05 | 7.702E-07 | 2.722E-08 |

**Table 4: Absolute differences between 11-term DTM and 11-term MDTM with RK4 solutions** $(\Delta t = 0.001)$.

|   | $|DTM_{0.01} - RK4_{0.001}|$ | | | $|MDTM_{0.01} - RK4_{0.001}|$ | | |
|---|---|---|---|---|---|---|
| $t$ | $\Delta T$ | $\Delta I$ | $\Delta V$ | $\Delta T$ | $\Delta I$ | $\Delta V$ |
| 0.04 | 8.554E-05 | 2.808E-10 | 5.846E-10 | 8.554E-05 | 2.808E-10 | 5.846E-10 |
| 0.27 | 5.868E-03 | 7.597E-08 | 1.615E-08 | 4.731E-05 | 6.696E-11 | 3.300E-10 |
| 0.88 | 2.441E-01 | 2.197E-06 | 8.952E-06 | 8.552E-05 | 3.735E-11 | 1.023E-10 |
| 1.60 | 7.744E+00 | 2.254E-05 | 5.067E-03 | 8.538E-05 | 6.653E-12 | 1.796E-11 |
| 5.00 | 2.513E+05 | 3.677E-01 | 9.109E+02 | 7.902E-05 | 9.648E-11 | 9.974E-11 |
| 8.80 | 5.954E+07 | 1.178E+01 | 3.266E+05 | 7.982E-05 | 4.704E-07 | 5.273E-07 |
| 10.00 | 2.083E+08 | 4.873E+02 | 1.222E+06 | 8.210E-05 | 3.357E-07 | 2.985E-06 |
| 12.40 | 1.728E+09 | 2.085E+03 | 1.117E+07 | 8.057E-05 | 3.948E-06 | 1.087E-06 |
| 14.80 | 9.908E+09 | 1.614E+04 | 6.840E+07 | 8.301E-05 | 2.005E-06 | 6.089E-07 |
| 16.00 | 2.142E+10 | 3.553E+04 | 1.517E+08 | 8.372E-05 | 1.408E-06 | 4.179E-07 |
| 18.00 | 6.875E+10 | 1.151E+05 | 5.047E+08 | 8.334E-05 | 6.671E-07 | 3.303E-08 |

The accuracy of the DTM and MDTM is investigated by comparing their solutions to the RK4 solution for the system parameters in Table 1 with the initial conditions $T(0) = 0.1$, $I(0) = 0$ and $V(0) = 0.1$. The RK4 with time steps $\Delta t = 0.01$ and $\Delta t = 0.001$ with the number of significant digits set to 16 is used. Table 3 and Table 4 presents the absolute errors between the 11-term DTM solutions and the 11-term MDTM solutions for the model of HIV infection of CD4$^+$T cells and the RK4 solutions with time steps $\Delta t = 0.01$ and $\Delta t = 0.001$ respectively.

In Table 3 and Table 4, we can observe that the DTM only gives valid result for $t \ll 1$. The MDTM solutions on the time step $\Delta t = 0.01$ agree with the RK4 solutions on the time step $\Delta t = 0.01$ to at least 4 decimal places, 5 decimal places and 6 decimal places for T, I and V respectively. Also, the MDTM solutions on the time step $\Delta t = 0.01$ agree with the RK4 solutions on a smaller time step $\Delta t = 0.001$ to at least 4 decimal places, 5 decimal places and 5 decimal places for T, I and V respectively. Hence, for the model of HIV infection of CD4$^+$T cells we observe that the MDTM solutions even agree with the RK4 solutions with a larger time step.

Figure 1-3 respectively presents the graphical representation of the concentration of CD4$^+$T cells, the concentration of infected CD4$^+$T cells by the HIV viruses and the concentration of free HIV virus particles. From Figure 2 which shows the concentration of infected CD4$^+$T cells, we observe that there is a "latent period" of about 750 days before the virus has a noticeable effect on the CD4$^+$T cell population. The concentration of CD4$^+$T cell population with and without HIV is shown in Figure 4. In the presence of HIV virus, when the CD4$^+$T cell count falls to 200 $cells/mm^3$ or below, an HIV patient is characterized to have developed AIDS.

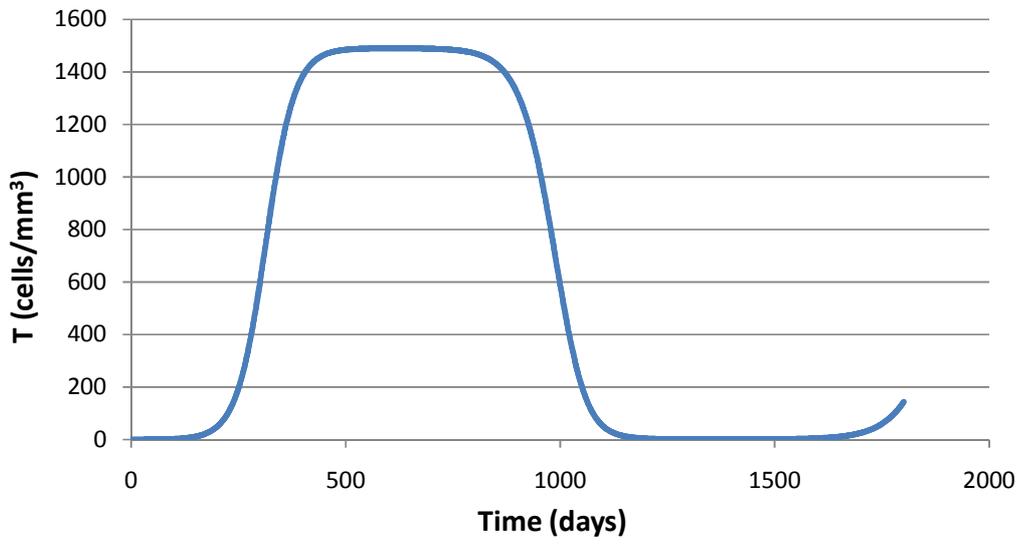

**Figure 1: Concentration of CD4⁺T cells with MDTM.**

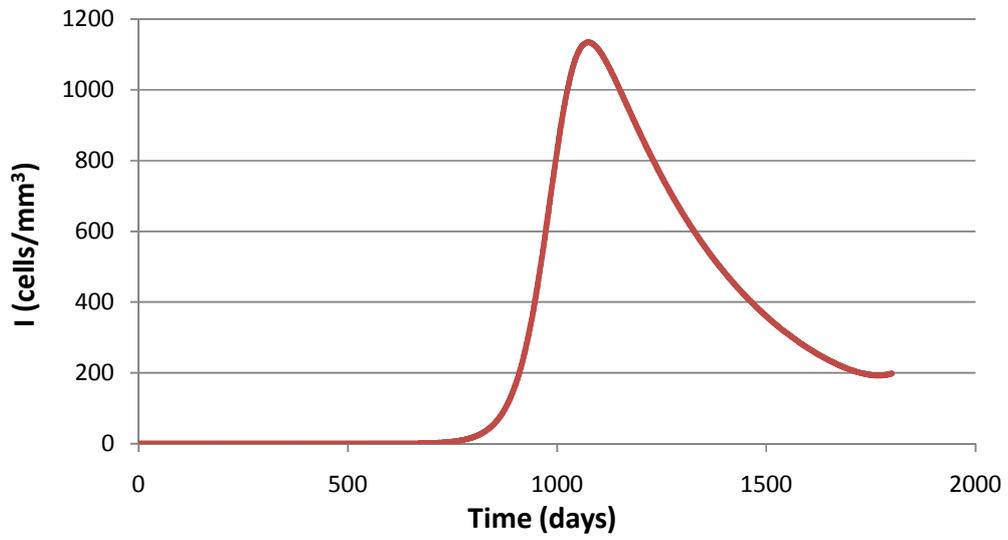

**Figure 2: Concentration of infected CD4⁺T cells by the HIV viruses with MDTM.**

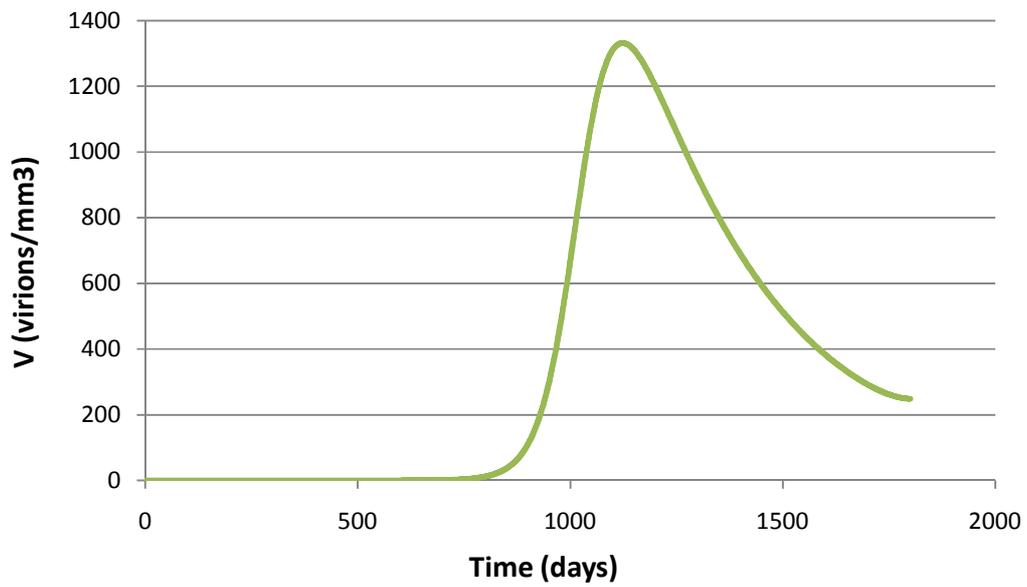

**Figure 3: Concentration of free HIV virus particles with MDTM.**

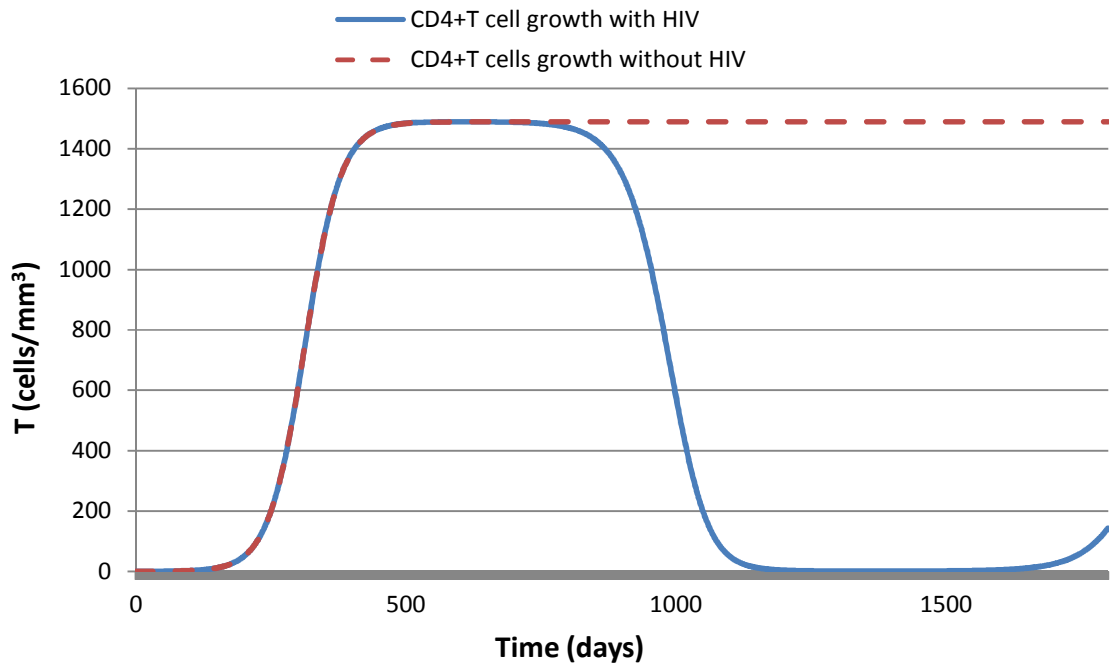

**Figure 4: Effect of HIV virus on growth of CD4$^+$T cells.**

## CONCLUSION

In this work, the DTM and MDTM were applied to solve model of HIV infection of CD4$^+$T cells. Comparisons were made between the DTM, MDTM and the fourth-order Runge-Kutta (RK4) method. The study shows that the method is a powerful and promising tool for solving this model and other coupled systems of differential equations.